\documentclass[aip,amsmath,amssymb,reprint]{revtex4-1}   
   
\usepackage{color}
\definecolor{indigo}{RGB}{0,0,120}
\usepackage[colorlinks=true, linkcolor=indigo, citecolor=blue, urlcolor=indigo]{hyperref}

\newcommand{\Bra}{\left \langle}
\newcommand{\Ket}{\right \rangle}
\newcommand{\bra}{\langle}
\newcommand{\ket}{\rangle}

\def\imply{\Rightarrow}

\def\fl{\noindent}

\newcommand{\beq}{\begin{equation}}
\newcommand{\eeq}{\end{equation}}
\newcommand{\beqs}{\begin{eqnarray}}
\newcommand{\eeqs}{\end{eqnarray}}

\newcommand{\half}{\frac{1}{2}}
\newcommand{\ov}[1]{\frac{1}{#1}}
\newcommand{\fr}[2]{\frac{#1}{#2}}

\def\del{\delta}
\def\eps{\epsilon} 
	
\def\sig{\sigma}

\def\vf{\varphi}		
\def\tht{\theta}	
\def\om{\omega}	
\def\Om{\Omega}  

\def\T{{\cal T}}
\def\V{{\cal V}}

\usepackage{subcaption}
\usepackage{ragged2e}
\DeclareCaptionJustification{justified}{\justifying}
\usepackage{graphicx}

\usepackage{calligra} 
\DeclareMathAlphabet{\mathcalligra}{T1}{calligra}{m}{n}
\DeclareFontShape{T1}{calligra}{m}{n}{<->s*[2.2]callig15}{}
\newcommand{\scripty}[1]{\ensuremath{\mathcalligra{#1}}}

\newcount\colveccount  
\newcommand*\colvec[1]{\global\colveccount#1  \begin{pmatrix} \colvecnext} \def\colvecnext#1{#1 \global\advance\colveccount-1
        \ifnum\colveccount>0 \\ \expandafter\colvecnext
        \else \end{pmatrix} \fi}

 \usepackage{soul}

\begin{document}



\title[Ergodicity, mixing and recurrence in the three rotor problem \hfill {\tt \href{https://arxiv.org/abs/1910.04455}{arXiv:1910.04455}}]{
Ergodicity, mixing and recurrence in the three rotor problem}

\author{Govind S. Krishnaswami}
\email{govind@cmi.ac.in}
\author{Himalaya Senapati}
\email{himalay@cmi.ac.in}
\affiliation{Physics Department, Chennai Mathematical Institute, SIPCOT IT Park, Siruseri 603103, India}

\date{10 April, 2020}

\begin{abstract}

Published in \href{https://doi.org/10.1063/1.5141067}{Chaos, {\bf 30} (4),  043112 (2020)} [Editor's pick]. \\

\fl In the classical three rotor problem, three equal point masses move on a circle subject to attractive cosine potentials of strength $g$. In the center of mass frame, energy $E$ is the only known conserved quantity. In earlier work [Krishnaswami and Senapati, arXiv:\href{https://arxiv.org/abs/1810.01317}{1810.01317}, Oct. 2018, arXiv:\href{https://arxiv.org/abs/1811.05807}{1811.05807}, Nov. 2018], an order-chaos-order transition was discovered in this system along with a band of global chaos for $5.33g \leq E \leq 5.6g$. Here, we provide numerical evidence for ergodicity and mixing in this band. The distributions of relative angles and angular momenta along generic trajectories are shown to approach the corresponding distributions over constant  energy hypersurfaces (weighted by the Liouville measure) as a power-law in time. Moreover, trajectories emanating from a small volume are shown to become uniformly distributed over constant energy hypersurfaces, indicating that the dynamics is mixing. Outside this band, ergodicity and mixing fail, though the  distributions of angular momenta over constant energy hypersurfaces show interesting phase transitions from Wignerian to bimodal with increasing energy. Finally, in the band of global chaos, the distribution of recurrence times to finite size cells is found to follow an exponential law with the mean recurrence time satisfying a scaling law involving an exponent consistent with global chaos and ergodicity.
\end{abstract}

\maketitle

\small

\tableofcontents

\vspace{1cm}

\normalsize

\begin{quotation}
The study of ergodicity on constant energy surfaces is an interesting question in dynamics. Geodesic flow on constant negative curvature Riemann surfaces provide a canonical well-studied example. In this paper, we find a new physical system, namely the classical three rotor problem, that displays global chaos and ergodicity in a band of energies. In this system, three equally massive point particles coupled via attractive cosine potentials move on a circle. Unlike in billiards, trajectories are smooth since the configuration space is a manifold without boundary. The quantum version is also of interest as it models chains of coupled Josephson junctions with rotor angles corresponding to superconducting phases. As the energy is increased, the system displays a fairly sharp transition to chaos followed by a band of global chaos and a subsequent gradual return to regularity. Here, we provide numerical evidence for ergodicity and mixing in this band of global chaos: time averages are shown to approach ensemble averages and trajectories from a small region are shown to spread out uniformly over constant energy hypersurfaces. Additional evidence for global chaos and ergodicity is provided through the investigation of recurrence time statistics. Interestingly, the system may be viewed as providing an example of ergodicity in geodesic flow on a torus with non-constant curvature of both signs.

\end{quotation}

\section{Introduction}
\label{s:introduction}

There are several  few degrees of freedom models that display global chaos as well as ergodicity and mixing. Geodesic flow on a constant negative curvature compact Riemann surface is a well-known example \cite{sinai-geodesic-flow,sinai-central-limit}. Ballistic motion on billiard tables of certain types including Sinai billiards \cite{sinai-billiard} and its generalization to the Lorentz gas \cite{lenci} provide other canonical examples. Kicked rotors and the corresponding Chirikov standard map \cite{chirikov} are also conjectured to display global chaos and ergodicity for certain sufficiently large parameter values \cite{stdmap-ergodicity}. In this paper, we study ergodicity and mixing in the classical three rotor problem which was shown\cite{gskhs-cnsd-3rotor,gskhs-3rotor} to display global chaos in a band of energies. This is in contrast with the model of three free but colliding masses moving on a circle where numerical investigations reported in Ref.~\cite{Rabouw-Ruijgrok} indicated a lack of ergodicity. An attractive feature of the three rotor system is that it offers the possibility of studying ergodicity in a continuous time autonomous Hamiltonian system of particles without boundaries or specular reflections. In particular, rotors can pass through each other without collisional singularities. This is physically reasonable since the three rotor problem is realized as the classical limit of a chain of coupled Josephson junctions \cite{sondhi-girvin} with the rotor angles corresponding to the superconducting phases associated to distinct segments between junctions. Interestingly, the center of mass dynamics of three rotors may also be regarded as geodesic flow on a 2-torus with non-constant curvature (of both signs) of an appropriate Jacobi-Maupertuis metric \cite{gskhs-3rotor}.

The statistics of recurrence times provides another window into chaotic dynamics \cite{kac, zaslavsky}. It is well-known that the distribution of recurrence times to small volumes in phase space approaches an exponential law for sufficiently mixing dynamics (e.g. Axiom-A systems \cite{hirata-axiomA} and some uniformly hyperbolic systems \cite{hirata-uniformly-hyperbolic}). Moreover, successive recurrence times are independently distributed so that the sequence of recurrence times is Poissonian. 

In this paper, we provide evidence for ergodicity in the band of global chaos by showing that numerically determined time averages approach the corresponding  ensemble averages. Evidence for mixing in the same band is obtained by showing that trajectories with a common energy from a small volume approach a uniform distribution on the energy hypersurface. Finally, we show that the distribution of recurrence times to finite size cells on such energy hypersurfaces follows an exponential law. Moreover, the mean recurrence time obeys a scaling law with exponent as expected from global chaos and ergodicity.

\begin{figure*}	
	\centering
	\begin{subfigure}[t]{4cm}
		\centering
		\includegraphics[width=4cm]{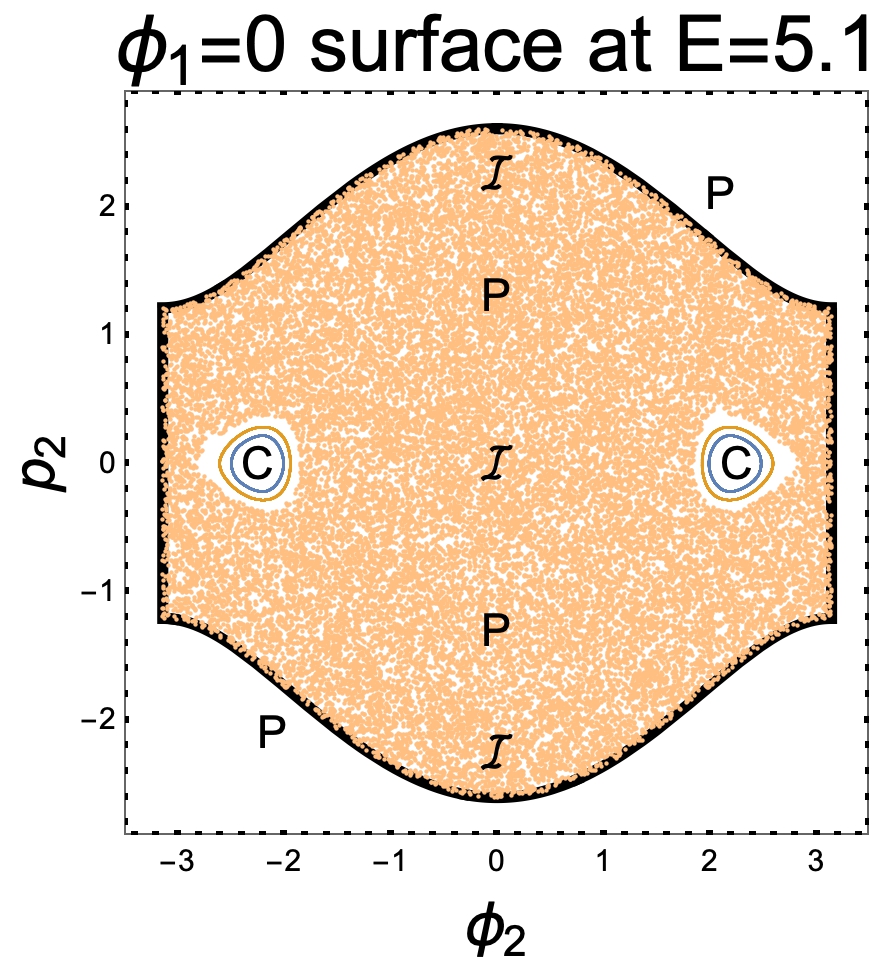}
	\end{subfigure}
\quad
	\begin{subfigure}[t]{4cm}
		\centering
		\includegraphics[width=4cm]{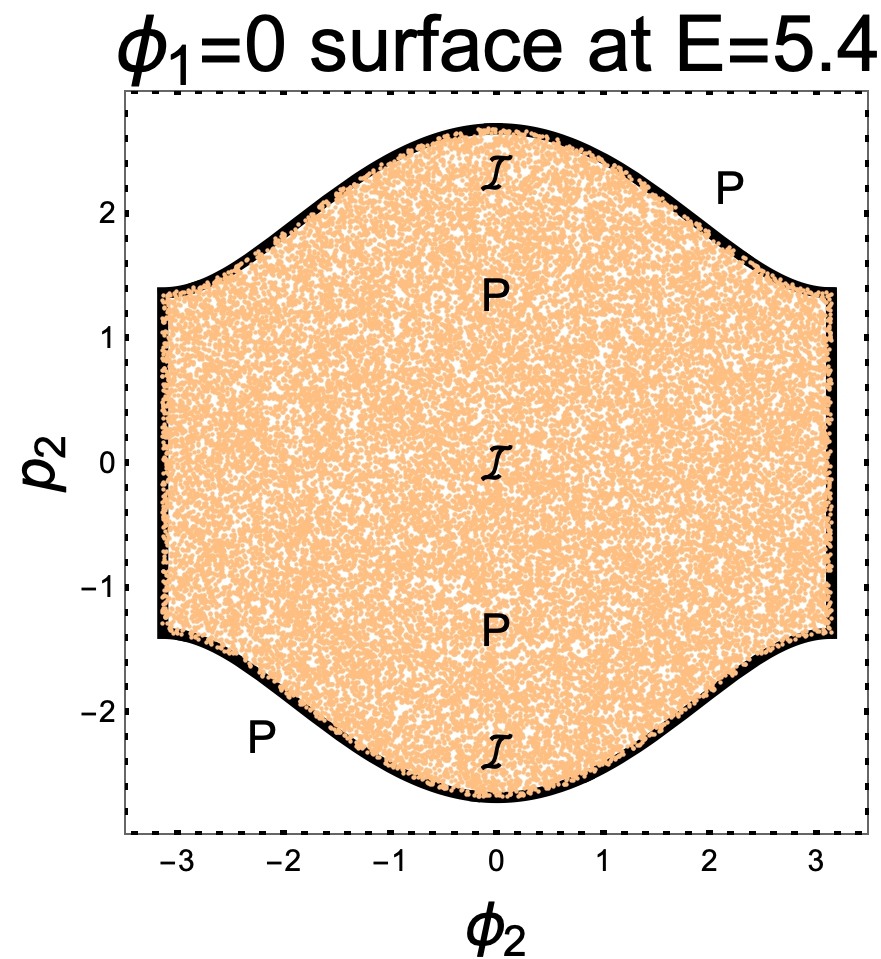}
	\end{subfigure}
\quad
	\begin{subfigure}[t]{4cm}
		\centering
		\includegraphics[width=4cm]{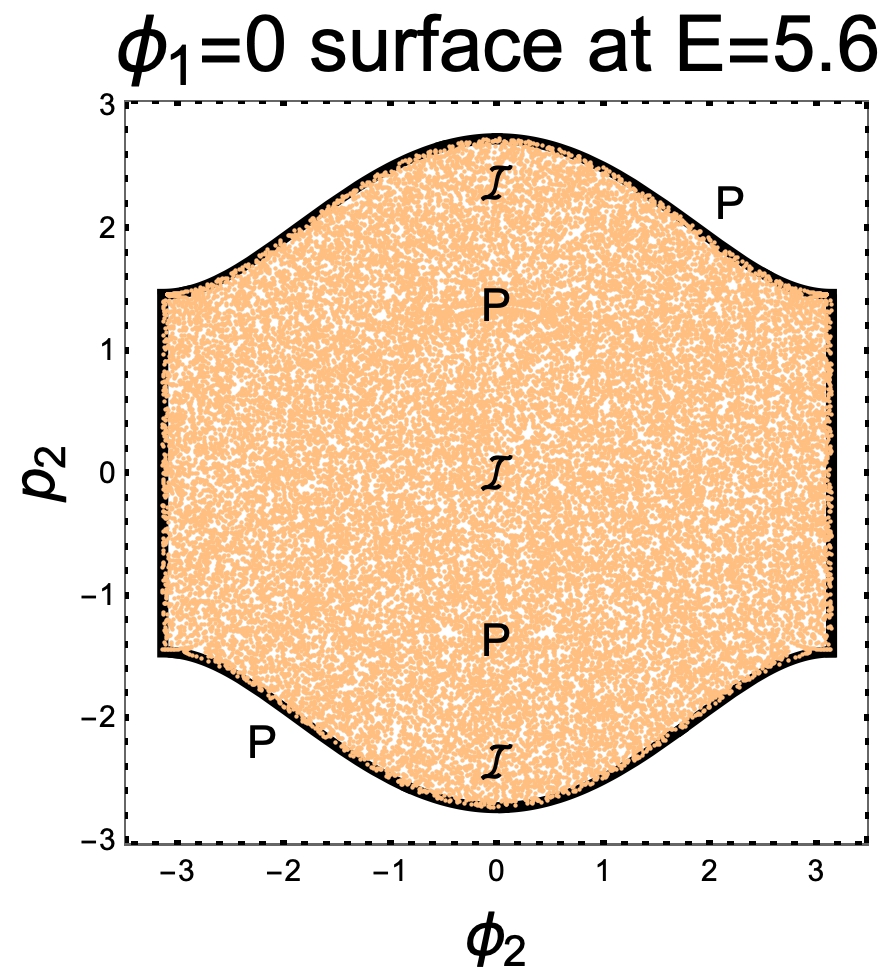}
	\end{subfigure}	
\quad
	\begin{subfigure}[t]{4cm}
		\centering
		\includegraphics[width=4cm]{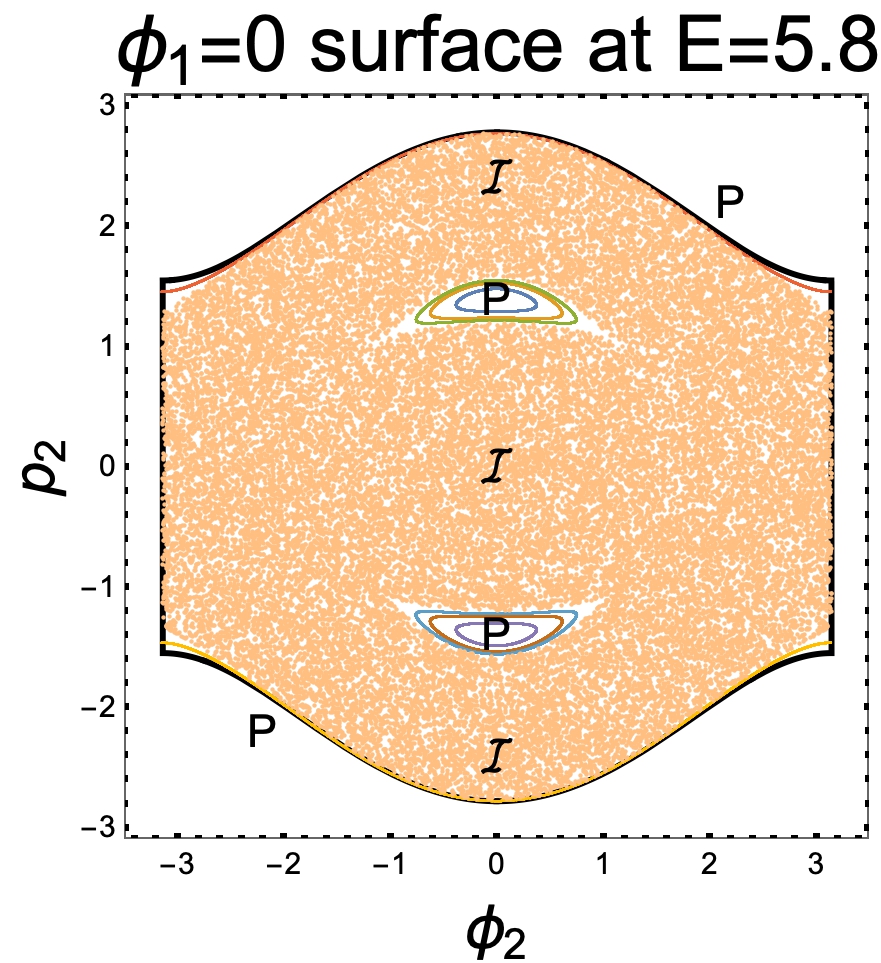}
	\end{subfigure}	
	\caption{\small Approach to the band of global chaos ($5.33g \leq E \leq 5.6g$) on the Poincar\'e surface $\vf_1 = 0$. The last elliptic islands to cease to exist (as $E \to 5.33g^-$) are around choreographies (C) and the first elliptic islands to open up (when $E$ exceeds $5.6g$) are around pendula (P) which also occur along the Hill boundary. Isosceles solutions intersect this surface at the points marked $\cal I$.}
	\label{f:poincare-section}
\end{figure*}

\section{Formulation and summary of results}
\label{s:formulation}

The classical dynamics of three rotors is governed by the Hamiltonian	
	\beq
	H = \sum_{i =1}^3  \left\{ \fr{\pi_i^2}{2 m r^2} + g [1- \cos\left(\tht_i-\tht_{i+1} \right) ] \right\}
	\eeq
where $\tht_4 \equiv \tht_1$, $\pi_i$ are the angular momenta conjugate to the rotor angles $\tht_i$ and $g \geq 0$ is the coupling strength. In center of mass and relative coordinates
	\beq
	\vf_0 = (\tht_1 + \tht_2 + \tht_3)/3, 
	\;\; \vf_1 = \tht_1 - \tht_2 \;\; \text{and} \;\;  \vf_2 = \tht_2 - \tht_3, 
	\eeq
the Lagrangian for {\it relative} motion is $L = \T - \V$ where 
	\beqs
	\T &=& \fr{1}{3} m r^2 \left[  \dot \vf_1^2 + \dot \vf_2^2 + \dot \vf_1\dot \vf_2 \right] \;\; \text{and} \cr
 \V &=& g \left[ 3 - \cos \vf_1  - \cos \vf_2  - \cos( \vf_1 +  \vf_2) \right].  
 	\label{e:lagrangian-phi-coords-3rotors}
 	\eeqs
The energy $E = \T + \V$ is the only known conserved quantity. The relative angles $\vf_{1,2}$ define periodic coordinates on a 2-torus configuration space and evolve independently of $\vf_0$:
	\beqs
	m r^2 \left(2 \ddot \vf_1 + \ddot \vf_2 \right) &=& - 3 g \left[ \sin \vf_1 + \sin ( \vf_1 + \vf_2) \right] \quad \text{and} \cr
	m r^2 \left(2 \ddot \vf_2 + \ddot \vf_1 \right) &=& - 3 g \left[ \sin \vf_2 + \sin ( \vf_1 + \vf_2) \right]. 
	\label{e:3rotors-cm-rel-coords-lagrangian-eom-2nd-order}
	\eeqs
These equations may be viewed as the equations for reparametrized geodesics on the configuration torus with respect to the Jacobi-Maupertuis metric line element \cite{gskhs-3rotor}
	\beq
	ds_{\rm JM}^2 = \fr{2 m r^2}{3} (E - \V) (d\vf_1^2 + d\vf_1 d\vf_2 + d\vf_2^2).
	\eeq
The system is integrable at $E=0$ and $E = \infty$ and displays a fairly sharp transition to chaos at $E \approx 4g$ and a more gradual return to regularity as $E \to \infty$ \cite{gskhs-3rotor}. This transition to stochasticity is manifested in a dramatic rise in the `fraction of chaos' ${\tt f}(E)$: the fraction of the area of the energetically allowed (Hill) region of the (for definiteness) `$\vf_1 = 0$' Poincar\'e surface occupied by chaotic sections. For instance, in Fig.~\ref{f:poincare-section}, $\tt f$ is the fractional area occupied by the orange colored chaotic sections. What is more, this dramatic rise in $\tt f$ (${\tt f} \approx 4\%$ at $E = 3.85g$ to ${\tt f} \approx 40\%$ at $E = 4.1g$) during the onset of chaos is accompanied by a spontaneous breaking of discrete symmetries of Poincar\'e sections. This transition is also associated with a change in the sign of the curvature of the Jacobi-Maupertuis metric defined on the $\vf_1$-$\vf_2$ configuration torus when $E$ exceeds $4g$ \cite{gskhs-3rotor,gskhs-3body}. Interestingly, the onset of chaos is also reflected in the stability of certain periodic solutions. In fact, in Ref.~\cite{gskhs-3rotor}, three families of periodic solutions of (\ref{e:3rotors-cm-rel-coords-lagrangian-eom-2nd-order}) were found: (a) {\it Pendula} where two of the rotors form a `molecule' with zero separation at all times, (b) {\it Isosceles breathers} where one rotor is always midway between the other two and (c) {\it Non-rotating choreographies} where all rotors are equally separated in time. Strikingly, the pendula display an accumulation of stability transitions as $E \to 4g^\pm$ coinciding with the onset of widespread chaos. 

Of special significance to the current work is the emergence of {\it global chaos} in the energy band $5.33g \leq E \leq 5.6g$. In this band, the fraction of chaos $\tt f$ on the `$\vf_1=0$' Poincar\'e surface attains the maximum value ${\tt f}\approx 1$ (see Fig.~\ref{f:poincare-section}). Intriguingly, the beginning of this band coincides with the divergence in the period of the non-rotating choreographies which additionally cease to exist above this energy. Similarly, the cessation of this band coincides with the energy at which pendula become stable.

Interestingly, in the above band, chaotic sections were also seen to occupy practically the whole of the energetically allowed portion of various other Poincar\'e surfaces (see Fig.~13 of Ref.~\cite{gskhs-3rotor}). This indicated that each chaotic trajectory explored the entire energy hypersurface justifying the term `global chaos'. Based on this and the roughly uniform manner in which chaotic sections filled up Poincar\'e surfaces, we conjectured that the dynamics is {\it ergodic} in this band of energies. The purpose of this paper is to investigate the dynamics in this band of global chaos and provide evidence for ergodicity and mixing, as well as study the distribution of recurrence times. 

We begin in \S \ref{s:ergodicity} by presenting evidence for ergodicity by showing that numerically determined time averages agree with ensemble averages.  In particular, we find the distributions of relative angles ($\vf_{1,2}$) and momenta ($p_{1,2}$) over constant energy hypersurfaces weighted by the Liouville measure. While the joint distribution function of $\vf_{1,2}$ is uniform on the Hill region of the configuration torus at all energies, the distribution of $p_1$ (and of $p_2$) shows interesting transitions from the Wigner semi-circular distribution when $E \ll g$ to a bimodal distribution for $E > 4.5g$ (see Fig.~\ref{f:collage-ensemble-time-avg-dist}).  In the band of global chaos, we find that distributions of $\vf_{1,2}$ and $p_{1,2}$ along generic (chaotic) trajectories are independent of the chosen trajectory and agree with the corresponding distributions over constant energy hypersurfaces, indicating ergodicity. This agreement fails for energies outside this band. In \S \ref{s:approach-to-ergodicity}, we investigate the rate of approach to ergodicity in the band of global chaos. We find that time averages such as $\bra \cos^2 \vf_1 \ket_{\rm t}$ and $\bra p_1^2 \ket_{\rm t}$ along a generic trajectory over the time interval $[0,T]$  approach the corresponding ensemble averages as a power law $\sim T^{-1/2}$ (see Fig. \ref{f:mean-vs-time}). This is expected of an ergodic system where correlations decay sufficiently fast in time  as shown in Appendix \ref{a:approach-to-ergodicity} (see also Ref.\cite{prl-dechant}).


In \S \ref{s:mixing}, we show that the dynamics is mixing (with respect to the Liouville measure) in the band of global chaos. This is done by showing that the histogram of number of trajectories in various cells partitioning the energy hypersurface approaches a distribution strongly peaked at the expected value with increasing time (see Fig. \ref{f:approach-to-mixing-in-time}). We also observe characteristic departures from mixing {\it even in chaotic regions} of the phase space at energies just outside this band (see Fig.~\ref{f:approach-to-mixing-in-energy}). 

In \S \ref{s:recurrenc-time-dist}, we study the distribution of recurrence times to a finite size cell \cite{altmann} in a given energy hypersurface. Within the band of global chaos, we find that the normalized distribution of recurrence times $\tau$ follows the exponential law $(1/\bar \tau) \exp(-\tau/\bar \tau)$ with possible deviations at small recurrence times (see Fig.~\ref{f:recurrence-time-distribution}). Though the mean recurrence/relaxation time $\bar \tau$ varies with the Liouville volume ${\rm v}$ of the cell, we find that it obeys the scaling law $\bar \tau \times {\rm v}^{2/3} = \tau^*$. This scaling law is similar to the ones discussed in  Refs.~\cite{balakrishnan-scaling-law, gao} with the scaling exponent $2/3$ consistent with global chaos and ergodicity. The rescaled mean recurrence time $\tau^*$ can vary with the location of the cell center, but does {\it not} vary significantly with energy in the band of global chaos. Finally, we demonstrate a loss of memory by showing that the gaps between successive recurrence times are uncorrelated.

\section{Ergodicity in the band of global chaos}
\label{s:ergodicity}

 In this section, we provide evidence for ergodicity in the band of global chaos ($5.33g \leq E \leq 5.6g$) by comparing distributions of $\vf_{1,2}$ and $p_{1,2}$ on constant energy hypersurfaces (weighted by the Liouville measure) with their distributions along generic numerically determined trajectories. For ergodicity, the distribution along a generic trajectory (over sufficiently long times) should be independent of initial condition and tend to the corresponding distribution over the energy hypersurface \cite{arnold-avez,gutzwiller-book}. We also examine the rate of approach to ergodicity in time and deviations from ergodicity outside the band of global chaos. Our numerical and analytical results, while indicative of ergodic behavior, are nonetheless not sufficient to establish it, since we examine only a restricted set of observables.

\subsection{Distributions along  trajectories and over energy hypersurfaces}
\label{s:ergodicity-time-ensemble-avgs}

\begin{figure*}	
\centering
	\includegraphics[width=18cm]{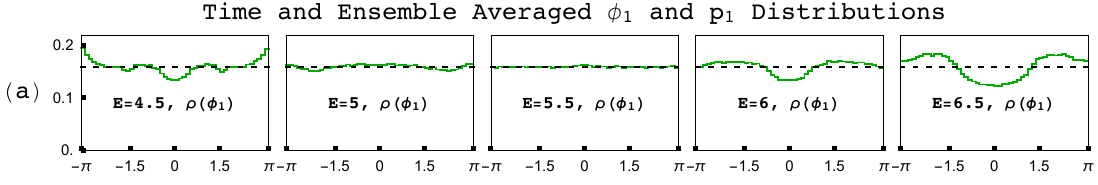}
	\includegraphics[width=18cm]{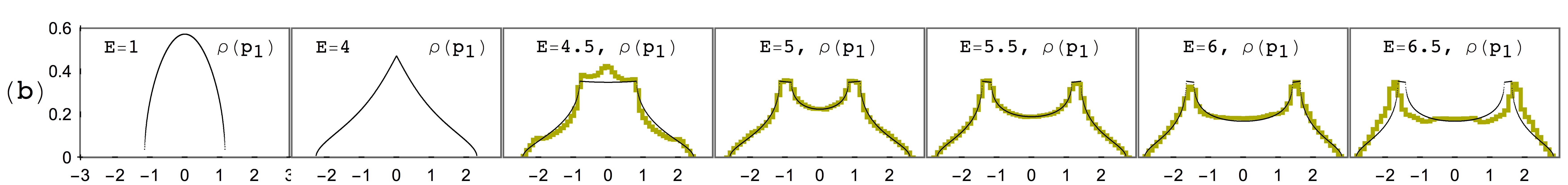}
	\caption{\small Distribution along generic trajectories (yellow, lighter) and distribution over constant energy hypersurface (black, darker) of (a) relative angle ($\vf_1$) and (b) relative momentum ($p_1$) for a range of increasing energies with $m = r = g = 1$. The horizontal axis is $\vf_1$ in (a) and $p_1$ in (b). Note that $\vf_1$ and $\vf_2$ have the same distributions as do $p_1$ and $p_2$. The  distribution along a generic (chaotic) trajectory is found to be insensitive to the IC chosen. The  momentum distribution over constant energy hypersurfaces transitions from a Wigner semi-circle to a bimodal distribution with increasing energy. The two distributions agree only in the band of global chaos $(5.33 \leq E \leq 5.6)$ consistent with ergodicity in this band.}
	\label{f:collage-ensemble-time-avg-dist}
\end{figure*}

\begin{figure*}	
	\centering
	\begin{subfigure}[t]{18cm}
		\centering
		\includegraphics[width=18cm]{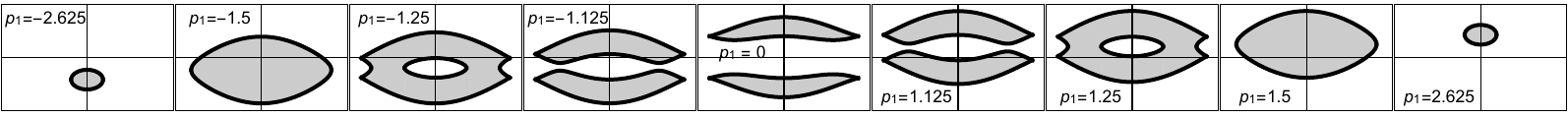}
	\end{subfigure}
	\caption{\small The energetically allowed portion (shaded gray) of the $\vf_2$-$p_2$ Poincar\'e surface for a sequence of increasing values of $p_1$ at $E = 5.5$ in the band of global chaos for $m = r = g = 1$.  On each plot, the horizontal axis is $\vf_2 \in [-\pi, \pi]$ and the vertical axis is $p_2 \in [-3, 3]$. The value of the distribution function $\rho_E(p_1)$ is the Liouville area of the shaded region. It is plausible that $\rho_E(p_1)$ is even and that as $p_1$ goes from $0$ to $p_{\rm max} = \sqrt{4 m r^2 E/3} \approx 2.71$, $\rho_E(p_1)$ initially increases from a non-zero local minimum, reaches a maximum and then drops to zero as shown in the $E = 5.5$ subfigure of Fig. \ref{f:collage-ensemble-time-avg-dist}b.}
	\label{f:collage-psec}
\end{figure*}

{\fl \bf Distribution along generic trajectories:} By the {\it distribution function} of a dynamical variable $F(p,\vf)$ (such as $p_1$ or $\vf_1$) along a {\it given trajectory} parametrized by time $t$, we mean
	\beq
	\varrho_F(f) = \lim_{T \to \infty} \ov{T} \int_{0}^T \del(F(p(t),\varphi(t)) - f) \: dt.
	\eeq
The time average of $F$ along the trajectory is then given by the first moment $\bra F \ket_{\rm t} = \int f \:\varrho_F(f) df$. In practice, to find the distribution of $F$, we numerically evolve a trajectory starting from a random initial condition (IC) and record the values $f$ of $F$ at equally spaced intervals of time (say, $\Delta t = .25$) up to $t_{\rm max} = 3 \times 10^5$ in units where $g = m = r = 1$. For such $t_{\rm max}$ and for energies in the globally chaotic band, we find that the histograms of recorded values  approach asymptotic distributions (see Fig.~\ref{f:collage-ensemble-time-avg-dist}) that are largely independent of the choice of $\Delta t$ and ICs.

{\fl \bf Distributions over energy hypersurfaces:} The ensemble average $\bra \cdot \ket_{\rm e}$ of a dynamical variable $F(p,\vf)$ at energy $E$ is defined with respect to the Liouville volume measure on phase space. Since $\vf_i$ and $p_j$ are canonically conjugate, we have
	\beqs
	&\bra F \ket_{\rm e}  = \ov{V_E} \int F \;\del(H - E) \; d\vf_1 \, d\vf_2 \, dp_1 \, dp_2 \cr
	 &\text{where} \quad	
	V_E = \int \del(H - E) d\vf_1 d\vf_2 dp_1 dp_2
	\label{e:ensemble-avg-distr-vol-egy-surf}
	\eeqs
is the volume of the $H = E$ energy hypersurface $M_E$. More generally, the {\it distribution} of $F(p,\vf)$ over the energy $E$ hypersurface weighted by the Liouville measure is the following phase space integral:
 	\beq
	\rho_{F,E}(f) = \ov{V_E} \int \del(F(p,\vf) - f) \del(H - E) d\vf_1 d\vf_2  dp_1 dp_2.
	\eeq
Loosely, it is like the Maxwell distribution of speeds in a gas. We will often omit the subscripts $F$ and/or $E$ when the observable and/or the energy are clear from the context. By definition, the above distribution is a probability density: $\int \rho(f) df = 1$. The ensemble average $\bra F \ket_{\rm e}$ is its first moment:
	\beq 
	\bra F \ket_{\rm e} = \int f \: \rho_{F,E}(f) \: df.
	\eeq
To find distributions over an energy hypersurface $M_E$, we need to integrate over it. For instance, to find the volume $V_E$ of the energy hypersurface, we observe that the Hamiltonian $H = \T + \V$ is quadratic in $p_2$ where
	\beqs
	\T &=&  \fr{p_1^2 + p_2^2 - p_1 p_2}{m r^2} \quad \text{and} \cr
	\V(\vf_1, \vf_2) &=& g \left[3 - \cos \vf_1  - \cos \vf_2  - \cos(\vf_1 + \vf_2) \right]. \quad
	\label{e:hamiltonian}
	\eeqs
Hence, we cover $M_E$ by two coordinate patches parametrized by $\vf_1, \vf_2$ and $p_1$ with
	\beq
	p_2^\pm = \half \left( p_1 \pm \sqrt{ 4 m r^2 (E-\V(\vf_1, \vf_2)) - 3 p_1^2} \right).
	\label{e:p2PM}
	\eeq
Using the factorization $H - E = (p_2 - p_2^+) (p_2 - p_2^-)$, we evaluate the integral over $p_2$ in Eq. (\ref{e:ensemble-avg-distr-vol-egy-surf}) to arrive at
	\beq
V_E =   \iint\displaylimits_{(\vf_1, \vf_2) \in {\cal H}_E} d\vf_1 \, d\vf_2 \int\displaylimits_{- p_{\rm max}}^{p_{\rm max}} \fr{dp_1}{(p_2^+ - p_2^-)}
	\eeq
where $p_{\rm max} = \sqrt{4mr^2 (E - \V)/{3}}$. Here, $\vf_{1,2}$ are restricted to lie in the Hill region ${\cal H}_E$ ($\V \leq E$). Interestingly, the integral over $p_1$ is independent of $\vf_1$ and $\vf_2$ as well as $E$ so that
	\beq
	\int\displaylimits_{- p_{\rm max}}^{p_{\rm max}} \fr{dp_1}{(p_2^+ - p_2^-)} = \fr{\pi}{\sqrt{3}} 
	\;\; \imply \;\;
	V_E = \fr{\pi}{\sqrt{3}} \times \text{Area}({\cal H}_E).
	\label{e:p1-integral}
	\eeq
Here, Area(${\cal H}_E$) is the area of the Hill region with respect to the measure $d\vf_1 d\vf_2$. It is a monotonically increasing function of $E$ and saturates at the value $4\pi^2$ for $E \geq 4.5$ when the Hill region includes the entire $\vf_1$-$\vf_2$ torus. We now derive formulae for distributions over energy hypersurfaces.

{\fl \bf Distribution of angles:} The {\it joint distribution function} of $\vf_1$ and $\vf_2$ is given by ($p_2^\pm$ are as in (\ref{e:p2PM}))
	\beqs
	\rho_E(\vf_1, \vf_2) &=& \ov{V_E} \int \del(H - E) \; dp_1 \, dp_2 \cr
	 &=& \ov{V_E} \int\displaylimits_{- p_{\rm max}}^{p_{\rm max}} \fr{dp_1}{(p_2^+ - p_2^-)} = \fr{\pi}{V_E\sqrt{3}},
	\eeqs
since from (\ref{e:p1-integral}), the integral over $p_1$ is $\pi/\sqrt{3}$ for all $E$ and $\vf_{1}$. In other words, ($\vf_1,\vf_2$) is uniformly distributed on the Hill region. Furthermore, for $E \geq 4.5$, the Hill region is the whole torus and $\rho_E(\vf_1, \vf_2) = 1/4\pi^2$. Thus, $\vf_1$ and $\vf_2$ are each uniformly distributed on $[0,2\pi]$ for $E \geq 4.5$. Fig.~\ref{f:collage-ensemble-time-avg-dist}a shows that the distributions of $\vf_1$ and $\vf_2$ along a trajectory with energy $E = 5.5$ in the band of global chaos agrees with this uniform phase space distribution (the fractional deviation is at most .2 \% across all angles).

{\fl \bf Distribution of momenta:} The momentum distribution functions turn out to be more intricate. Due to the $1\leftrightarrow 2$ symmetry of the Hamiltonian (\ref{e:hamiltonian}), the 1-particle momentum distribution functions $\rho_E(p_1)$ and $\rho_E(p_2)$ are equal and given by the marginal distribution
	\beqs
	\rho_E(p_1) &=& \ov{V_E} \int \del(H - E) \; d\vf_1 \, d\vf_2 \, dp_2 \cr
	&=& \ov{V_E} \iint\displaylimits_{(\vf_1, \vf_2) \in {\cal H}_{E,p_1}} \fr{d\vf_1 \, d\vf_2}{p_2^+ - p_2^-}.
	\eeqs
Here, ${\cal H}_{E,p_1}$ is the portion of the $\vf_1$-$\vf_2$ torus allowed for the given values of $E$ and $p_1$. Since $p_2^\pm$ must be real, from (\ref{e:p2PM}) we see that $4mr^2 (E - \V) - 3 p_1^2 \geq 0$ or $\V \leq E - 3 p_1^2/4 mr^2$. Thus, $\vf_1$ and $\vf_2$ must lie in the Hill region for the modified energy $E' = E - 3 p_1^2/4 mr^2$. For this Hill region to be non-empty, we must have $E' \geq 0$. Thus, the distribution function $\rho_E(p_1)$ is supported on the interval $[-\sqrt{4 m r^2 E/3}, \sqrt{4 m r^2 E/3}]$ and is given by 
	\beq
	\rho_E(p_1) = \ov{V_E} \iint_{{\cal H}_{E'}} \fr{d\vf_1 \, d\vf_2}{ \sqrt{4 m r^2 (E'(p_1) - \V)}}.
	\label{e:rhoe-p1-general}
	\eeq
On account of $E'(p_1)$ being even, $\rho_E(p_1) = \rho_E(-p_1)$. Upon going to Jacobi coordinates $\vf_\pm = (\vf_1 \pm \vf_2)/2$, the integral over $\vf_-$ can be expressed in terms of an incomplete elliptic integral of the first kind. Though the resulting formulae are lengthy in general, for low energies $\rho_E(p_1)$ turns out to be the Wigner semi-circular distribution (see Fig. \ref{f:collage-ensemble-time-avg-dist}b). Indeed, upon going to Jacobi coordinates and using the quadratic approximation for the potential $\V_{\rm low} = 3 g \vf_+^2 + g \vf_-^2$, we find that at low energies, the Hill region ${\cal H}_{E'}$ is the elliptical disk $ 3 g \vf_+^2 + g \vf_-^2 \leq E'(p_1)$. Thus, 
	\beq
	V_E = \fr{\pi}{\sqrt{3}} \times {\rm Area}({\cal H}_E) = \fr{2 \pi^2 E}{3 g} \quad \text{for} \quad E \ll g
	\eeq
leading to the Wigner semi-circular distribution
	\beqs
	\rho_E (p_1) &=& \ov{V_E} \iint_{{\cal H}_{E'}} \fr{2 d\vf_+ d\vf_-}{\sqrt{4 m r^2(E'(p_1) - \V_{\rm low})}} \cr
	&=& \fr{3}{2 \pi m r^2 E} \sqrt{\fr{4}{3} m r^2 E - p_1^2} \quad \text{for} \quad E \ll g.
	\eeqs
For larger $E$, we perform the integral (\ref{e:rhoe-p1-general}) numerically. Fig. \ref{f:collage-ensemble-time-avg-dist}b shows that the distribution goes from being semi-circular to bimodal as $E$ crosses $4g$. Loosely, $\rho_E(p_1)$ is the analogue of the Maxwell distribution for the relative momenta of the three rotor problem. Fig. \ref{f:collage-psec} provides a qualitative explanation of the bimodal shape of $\rho_E(p_1)$ for an energy in the band of global chaos. Fig. \ref{f:collage-ensemble-time-avg-dist}b shows that the distribution of $p_1$ along a generic trajectory closely matches its distribution $\rho_E(p_1)$ over the constant energy hypersurface in the band of global chaos $(5.33 \leq E \leq 5.6)$ but deviates at other energies, providing evidence for ergodic behavior in this band.

\subsection{Approach to ergodicity}
\label{s:approach-to-ergodicity}

\begin{figure*}	
	\centering
	\begin{subfigure}[t]{5.3cm}
		\centering
		\includegraphics[width=5.3cm]{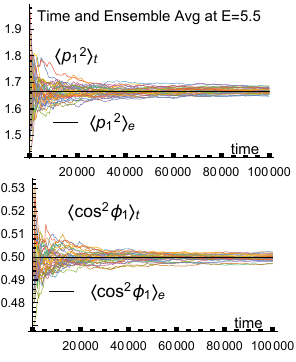}
		\caption{\small }
	\end{subfigure}
\quad
	\begin{subfigure}[t]{10cm}
		\centering
		\includegraphics[width=10cm]{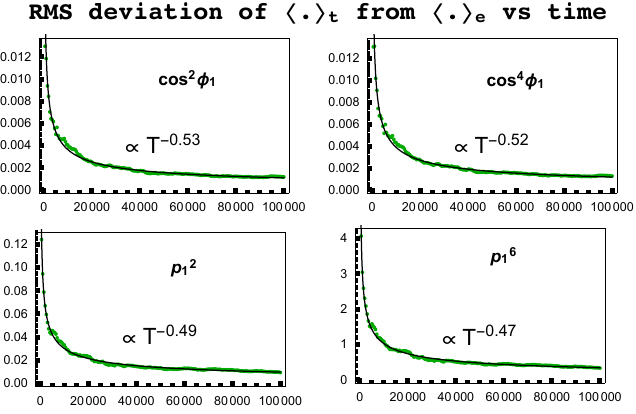}
		\caption{\small }
	\end{subfigure}	
	\caption{\small (a) Time averages $\bra p_1^2 \ket_{\rm t}$ and $\bra \cos^2 \vf_1 \ket_{\rm t}$ as a function of time $T$ for 35 randomly chosen trajectories at $E = 5.5$. They are seen to approach the corresponding ensemble averages ($\bra \cdot \ket_{\rm e}$ indicated by thick black lines) as time grows. (b) Root mean square deviation (over 35 chaotic initial conditions) of time averages from the corresponding ensemble average as a function of time $T$ for $E = 5.5$ in the band of global chaos for the observables $\cos^2 \vf_1$, $\cos^4 \vf_1$, $p_1^2$ and $p_1^6$. The fits show a $T^{-1/2}$ approach to ergodicity.}
	\label{f:mean-vs-time}
\end{figure*}

To examine the rate of approach to ergodicity for energies in the band of global chaos, we compare ensemble averages of variables such as $\cos^2 \vf_1$ and $p_1^2$ with their time averages over increasingly long times.

{\fl \bf Ensemble average:} The ensemble average $\bra \cdot \ket_{\rm e}$ of a variable $F$ at energy $E$ defined in (\ref{e:ensemble-avg-distr-vol-egy-surf}) reduces to 
	\beqs
	\bra F \ket_{\rm e} &=& \ov{V_E} \iint\displaylimits_{(\vf_1, \vf_2) \in {\cal H}_E} d\vf_1 \, d\vf_2 \int\displaylimits_{- p_{\rm max}}^{p_{\rm max}} \fr{F_{\rm sum} \; dp_1}{2(p_2^+ - p_2^-)} \cr
	\text{with} \;\; &F_{\rm sum}& =  F(\vf_1, \vf_2, p_1, p_2^+) + F(\vf_1, \vf_2, p_1, p_2^-)
	\eeqs
upon using the factorization $H - E = (p_2 - p_2^+) (p_2 - p_2^-)$ to evaluate the integral over $p_2$. Since for $E \geq 4.5$, $\vf_1$ and $\vf_2$ are independently uniformly distributed on $[0,2\pi]$, we have
	\beq
	\bra \cos^{m} \vf_1 \cos^{n} \vf_2 \ket_{\rm e} = \bra \cos^{m} \vf_1 \ket_{\rm e} \bra \cos^{n} \vf_2 \ket_{\rm e}
	\eeq
with $\bra \cos^{2n} \vf_1 \ket_{\rm e} = \fr{(2n)!}{2^{2n} (n!)^2}$ and the odd moments vanishing. Remarkably, the phase space averages of momentum observables are also exactly calculable for $E \geq 4.5$:
	\beqs
	\bra p_1^2 \ket_{\rm e} &=& \fr{2}{3} E - 2, \;\;\bra p_1^4 \ket_{\rm e} = \fr{2}{3} E^2 - 4 E + 7, \cr
	\bra p_1^2 p_2^2 \ket_{\rm e} &=& \fr{1}{3}E^2 - 2 E + \fr{7}{2} 
	\;\; \text{and} \cr
	\bra p_1^6 \ket_{\rm e} &=&\fr{20}{27}E^3  - \fr{20}{3}E^2 + \fr{70}{3}E - \fr{260}{9}.
	\eeqs
Though we restrict to $E \geq 4.5$ to obtain simple formulae for ensemble averages, this includes the band of global chaos $5.33 \leq E \leq 5.6$ where {\it alone} we can expect ergodic behavior. 

To compare with time averages, for each energy, we pick $N_{\rm traj} = 35$ random ICs (on the $\vf_1 = 0$ surface) and evolve them forward. As Fig. \ref{f:mean-vs-time}a indicates, though the time averages ($\ov{T} \int_0^T F \: dt$) display significant fluctuations at early times, they have approached their asymptotic values by $T = 10^5$. To estimate the rate of approach to ergodicity, we compute the root mean square deviation $\sigma(T)$ of the time average from the ensemble average as a function of time:
	\beqs
	\sigma^2(T) &=& \ov{N_{\rm traj}} \sum_a \left( \bra F \ket_{t, a}(T)  - \bra F \ket_{\rm e} \right)^2 \cr
	\text{where} &\quad& \bra F \ket_{t, a}(T) = \ov{T} \int_0^T F(t'_a) \: dt'_a
	\eeqs
is the time average over the $a^{\rm th}$ trajectory. Fig. \ref{f:mean-vs-time}b shows that for several variables $F = \cos^2 \vf_1, p_1^2$ etc., the mean square deviation decays roughly as the reciprocal of time, $\sig \sim 1/\sqrt{T}$, as expected of an ergodic system where correlations decay sufficiently fast  (see Appendix \ref{a:approach-to-ergodicity} and Ref.~\cite{prl-dechant}).

\begin{figure}	
	\centering
	\begin{subfigure}[t]{7cm}
		\centering
		\includegraphics[width=7cm]{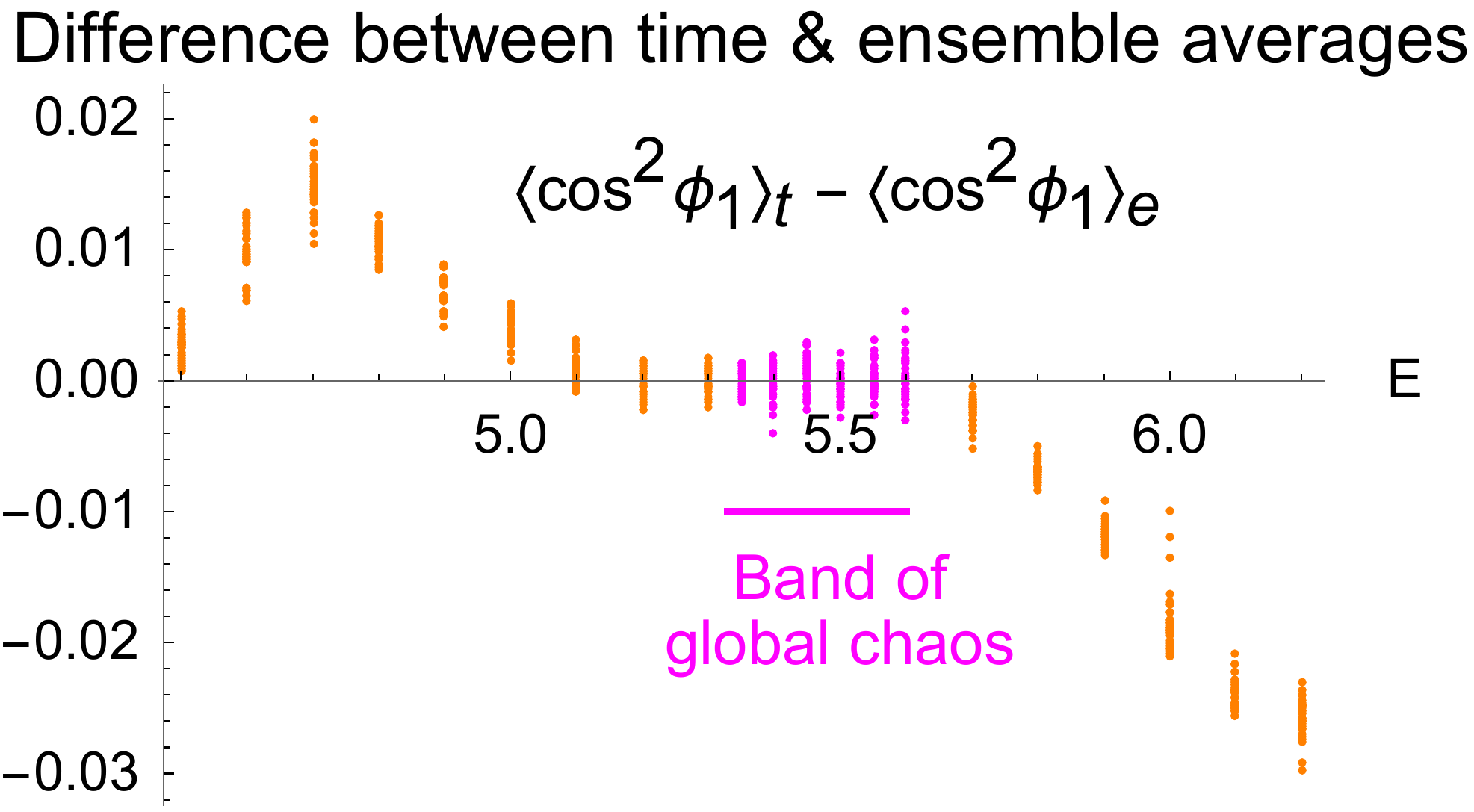}
	\end{subfigure}
	\begin{subfigure}[t]{7cm}
		\centering
		\includegraphics[width=7cm]{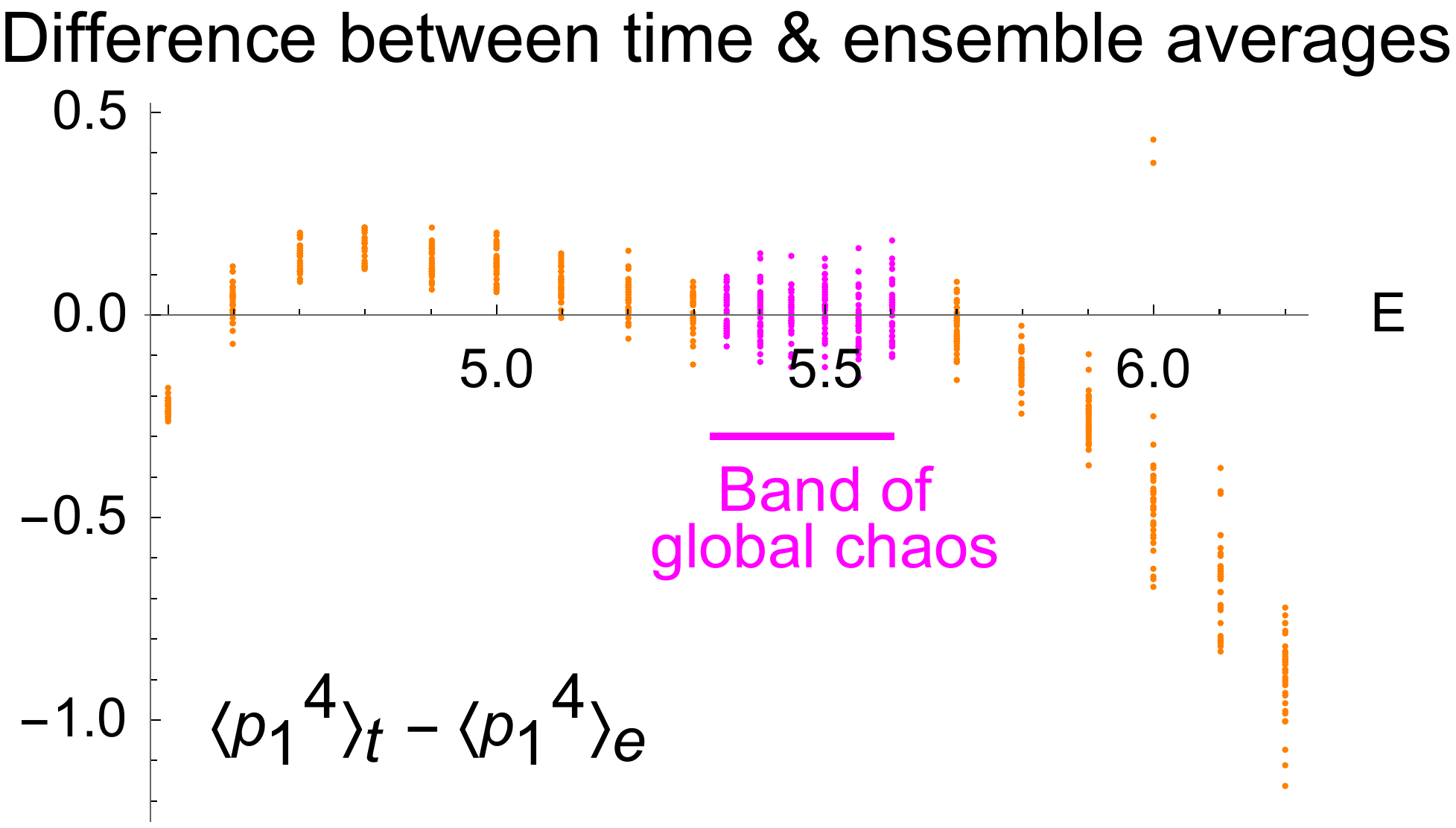}
	\end{subfigure}	
	\caption{ \small  Difference between time averages $\bra \cdot \ket_{\rm t}$ over a time $T = 10^5$ (for 35 randomly chosen chaotic trajectories) and ensemble average $\bra \cdot \ket_{\rm e}$ for $\cos^2 \vf_1$ and $p_1^2$ indicating ergodicity in the band of global chaos $5.33 \leq E \leq 5.6$ (magenta) and discernible departures outside this band (orange). The spread in $\bra \cdot \ket_{\rm t} - \bra \cdot \ket_{\rm e}$  at a fixed energy is due to the finiteness of $T$. However, this spread is small compared to  the average values $\bra \cos^2 \vf_1 \ket_{\rm e} = .5$ and $\bra p_1^4 \ket_{\rm e} = 2E^2/3 - 4E + 7$ demonstrating that time averages over distinct chaotic trajectories converge to a common value. Note that the spread in $\bra p_1^4 \ket_{\rm t} - \bra p_1^4 \ket_{\rm e}$ increases with $E$ as the average values themselves increase with $E$.}
	\label{f:mean-vs-egy}
\end{figure}

 Finally, we examine the approach to ergodicity as the energy approaches the band of global chaos $5.3 \lesssim E \lesssim 5.6$. To this end, we compare the  ensemble averages of a few variables with their time averages for 35 randomly chosen {\it chaotic trajectories} over a range of energies. Fig. \ref{f:mean-vs-egy} shows that the time averages of $\cos^2 \vf_1$ and $p_1^2$ agree reasonably well with their ensemble averages in the band of global chaos. At lower and higher energies, there are discernible deviations from the ensemble averages, showing ergodicity breaking. (a) For $E$ slightly outside the band of global chaos, we find that there is a single chaotic region (see Fig. \ref{f:poincare-section}), and time averages along trajectories from this region converge to a common value which however differs from the ensemble average over the whole energy hypersurface (see Fig. \ref{f:mean-vs-egy}). (b) At energies significantly outside the band of global chaos, there can be several distinct chaotic regions (see Fig. 11e of Ref. \cite{gskhs-3rotor}). We find that time averages of an observable along chaotic trajectories from these distinct regions generally converge to different values, none of which typically agrees with the ensemble average over the whole energy hypersurface.


\section{Mixing in the band of global chaos}
\label{s:mixing}

\begin{figure*}	
	\centering
	\begin{subfigure}[t]{8cm}
		\centering
		\includegraphics[width=8cm]{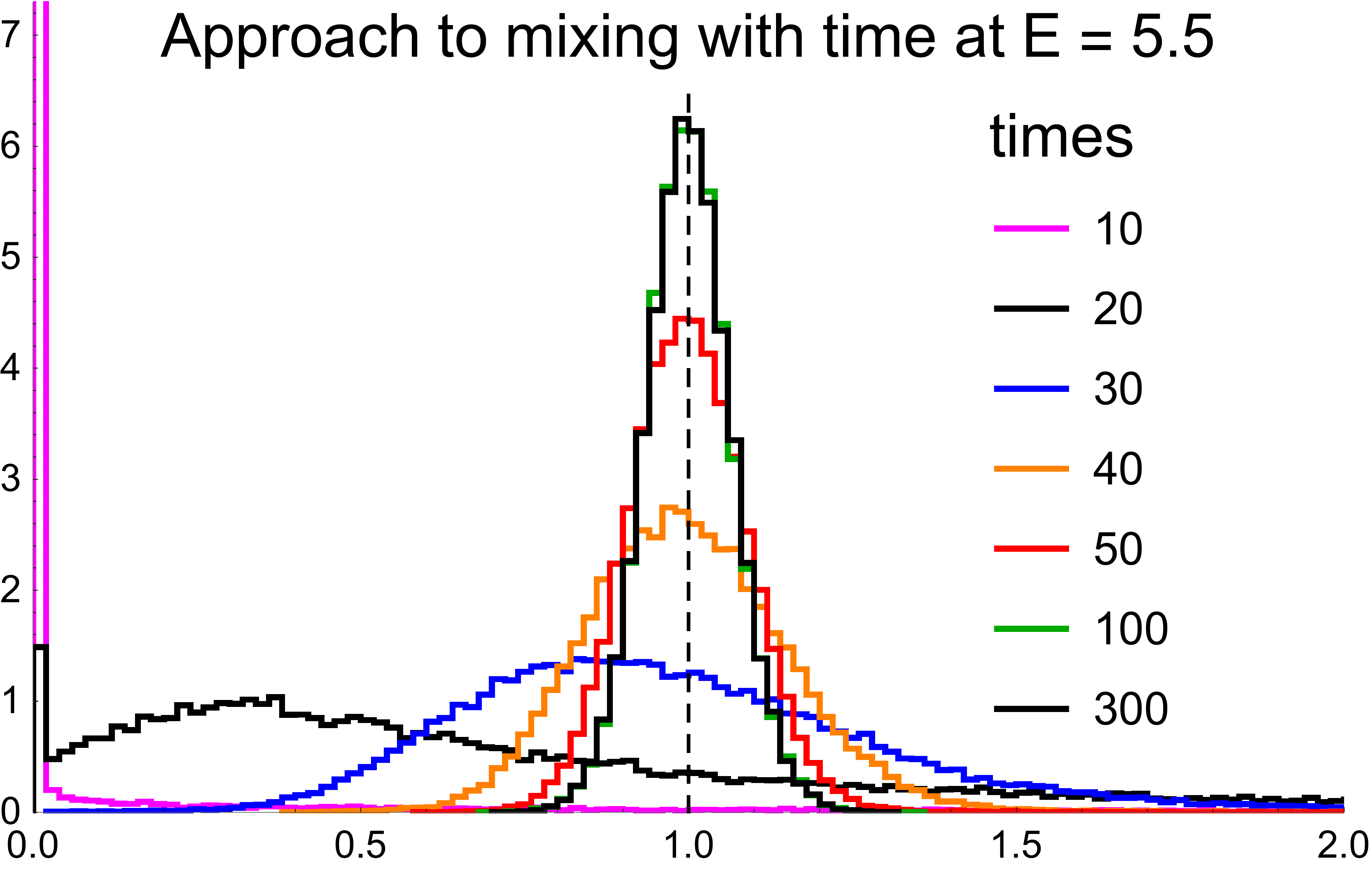}
		\caption{ \small  }
		\label{f:approach-to-mixing-in-time}
	\end{subfigure}
\quad
	\begin{subfigure}[t]{7.3cm}
		\centering
		\includegraphics[width=7.3cm]{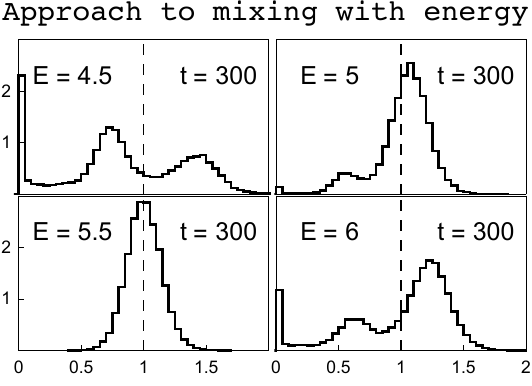}
		\caption{ \small  }
		\label{f:approach-to-mixing-in-energy}
	\end{subfigure}		
	\caption{ \small Histograms of number of trajectories $n_i(t)$ in each cell $i$ of an energy hypersurface. To facilitate comparison across energies and numbers of ICs considered, the histograms of $\tilde n_i(t) = (n_i(t) V_E) / (\mu_i N)$ (see Eq. \ref{e:mean-counts-mixing}) are displayed. For the flow to be mixing, the histograms should strongly peak around $\tilde n_i(t) = 1$. Fig. (a) shows the approach to mixing in time at an energy $E = 5.5$ in the band of global chaos. The histogram is seen to migrate from peaking at zero to 1 with advancing time. Fig. (b) shows these histograms at reasonably late times ($t=300$) showing how the flow becomes mixing as we approach the band of global chaos (represented here by $E = 5.5$).}
	\label{f:approach-to-mixing}
\end{figure*}

\begin{figure}	
	\centering
		\includegraphics[width=7cm]{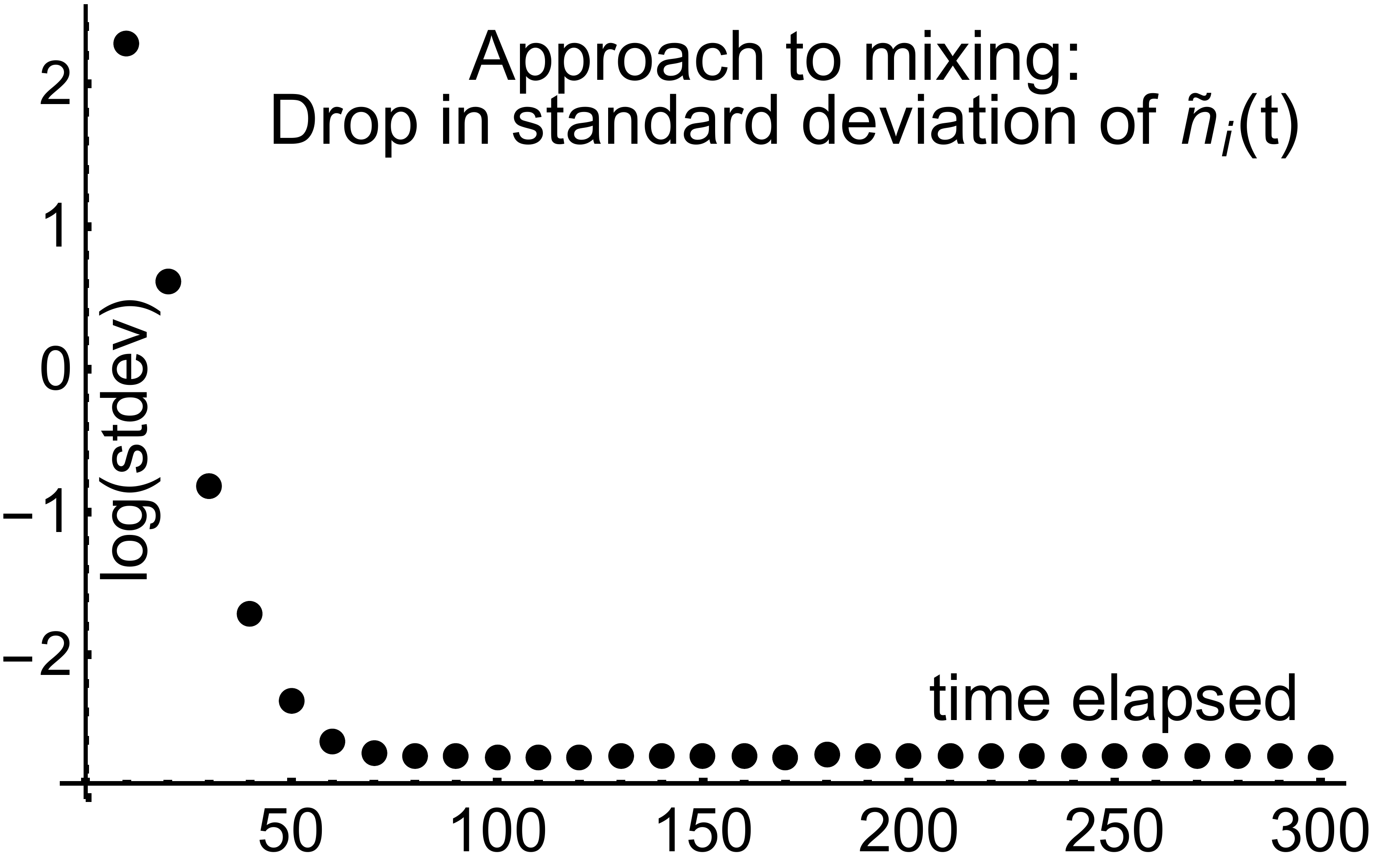}
	\caption{\small Drop with time of the standard deviation of the distribution (see Fig.~\ref{f:approach-to-mixing-in-time}) of the scaled number of trajectories $\tilde n_i(t)$ in each cell of the energy $E=5.5$ hypersurface.  The latter is partitioned into $N_{\rm cells} \approx 4 \times 10^4$ cells and $N = 1.3 \times 10^7$ trajectories have been considered. The plot shows that the standard deviation has dropped to $0.066$ at $t = 300$. This is close to the expected standard deviation $0.055$ if the $N$ trajectories were distributed uniformly among the $N_{\rm cells}$ cells at the instant considered.}
	\label{f:stdev-of-ntilde}
\end{figure}

\begin{figure}	
	\centering
	\begin{subfigure}[t]{4cm}
		\centering
		\includegraphics[width=4cm]{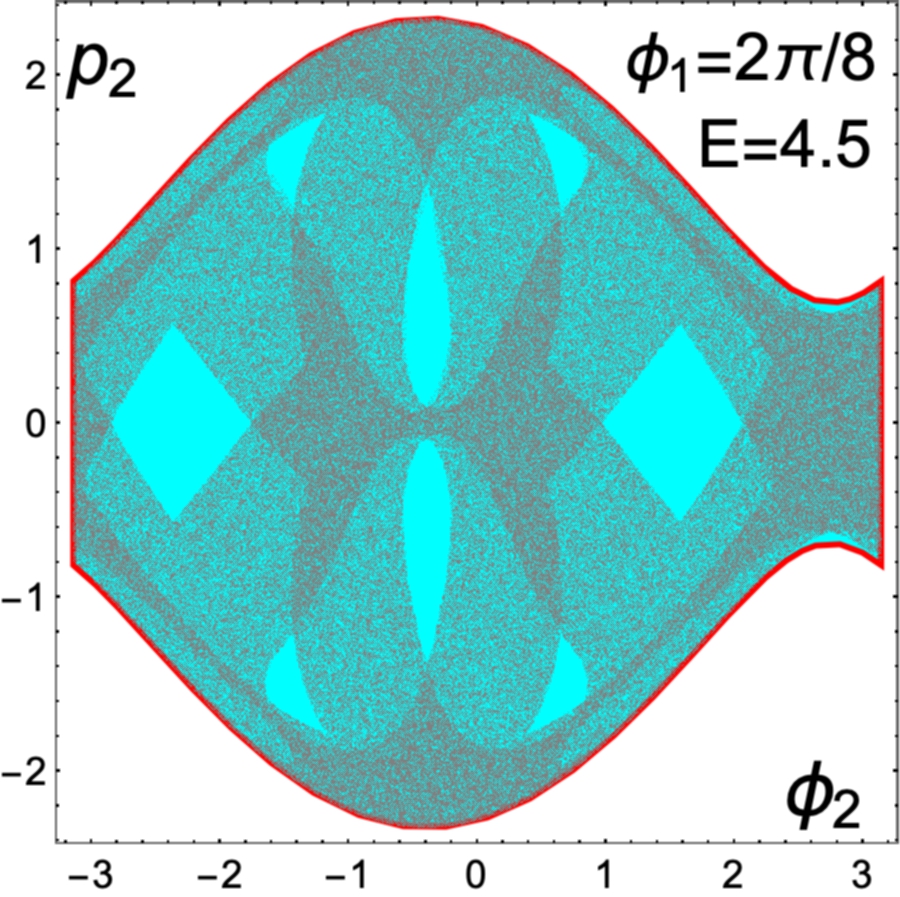}
	\end{subfigure}
\quad
	\begin{subfigure}[t]{4cm}
		\centering
		\includegraphics[width=4cm]{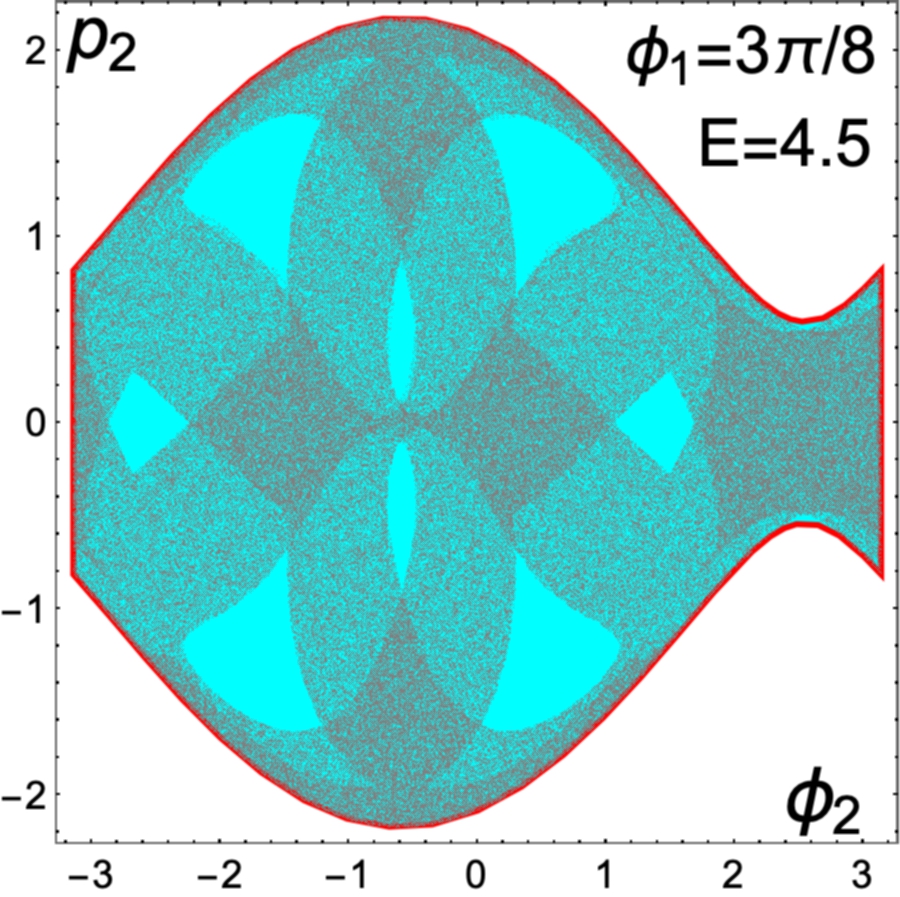}
	\end{subfigure}
\caption{\small Two distinct densities (shaded dark and light) of points (from trajectories for $0 \leq t \leq 10^5$) on chaotic sections of Poincar\'e surfaces at $E = 4.5$ corresponding to the two non-zero peaks in the histogram of Fig.~\ref{f:approach-to-mixing-in-energy} showing characteristic departure from mixing. The unshaded regions are energetically allowed but are not visited by these chaotic trajectories and correspond to the peak around zero in the same histogram.}
\label{f:psec-outside-globalchaos}
\end{figure}

In \S \ref{s:ergodicity}, we provided numerical evidence for ergodicity in the 3 rotor problem for energies in the band of global chaos. We now investigate whether the dynamics is mixing in this regime. A flow $\phi_t$ on the energy hypersurface $M_E$ of the phase space is said to be strongly mixing if for all subsets $A, B \subseteq M_E$ with positive measures ($\mu(A)>0$ and $\mu(B) >0$), we have
	\beq
	\lim_{t \to \infty} \mu(\phi_t(B) \cap A) = \mu(B) \times \mu(A)/\mu(M_E)
	\eeq
where $\mu$ is the Liouville volume measure on $M_E$ \cite{gutzwiller-book,arnold-avez}. To numerically examine whether the dynamics of three-rotors is mixing in the band of global chaos, we work in units where $m=r=g=1$ and consider a large number $N$ ($= 1.3 \times 10^7$) of random ICs with energy $E$ in a small initial region of phase space (e.g., $|\vf_{1,2}|, |p_1| < .05$ with $p_2 = p_2^+$ (\ref{e:p2PM}) determined by $E$). The trajectories are numerically evolved forward in time and their locations recorded at discrete time intervals (e.g., $t = 10$, 20, $\cdots$, 300). If the dynamics is mixing, then in the limit $N \to \infty$ and $t \to \infty$, the number of trajectories located at time $t$ in a Liouville volume $V$ must equal $N V/V_E$ where $V_E$ is the Liouville volume of the energy hypersurface. Poincar\'e sections (see Fig.~\ref{f:poincare-section}) as well as investigations of ergodicity in \S \ref{s:ergodicity} rule out the possibility of mixing for energies outside the regime of global chaos. Thus, we restrict to $5.33 \leq E \leq 5.6$ where $V_E = {4\pi^3}/\sqrt{3}$, a formula that holds for any $E \geq 4.5$ (\ref{e:p1-integral}). Now, for convenience, we divide the 3d energy hypersurface into cuboid-shaped cells of equal geometric volume $V^g$. The Liouville volumes of these cells are not equal, so we denote by $\mu_i$ the Liouville volume of the $i^{\rm th}$ cell. In practice, we take cells of linear dimensions ${2\pi}/{d}$ each in $\vf_1$ and $\vf_2$ and ${2p_1^{\rm max}}/{d}$ in $p_1$ where $d = 40$ is the number of subdivisions and $p_1^{\rm max}$ the maximal value of $p_1$ corresponding to energy $E$. Though we compute $\mu_i$ exactly, it is approximately $V^g \times$ the Liouville density at the center of the $i^{\rm th}$ cell:
	\beq
	\mu_i \approx \ov{2(p_2^+ - p_2^-)} \times \fr{2\pi}{d} \times \fr{2\pi}{d} \times \fr{2p_1^{\rm max}}{d}
	\eeq 
where $p_2^\pm$ (\ref{e:p2PM}) are evaluated at the center of the cell. Cells that lie outside or straddle the boundary of the energy hypersurface are not considered. At various times, we record the instantaneous locations of the trajectories and count the number $n_i(t)$ of trajectories that lie in the cell $i$. If the dynamics is mixing, the number of trajectories in the $i^{\rm th}$ cell should be
	\beq
	n_i = N \times \fr{\mu_i}{V_E}.
	\label{e:mean-counts-mixing}
	\eeq 
To test the mixing hypothesis and rate of approach, we plot in Fig.~\ref{f:approach-to-mixing} at various times $t = 10, 20, \cdots, 300$, a histogram of $n_i(t)$. To be more precise, we plot a histogram of $\tilde n_i(t) = n_i(t) V_E / (\mu_i N)$ so that the expected mean is 1, to facilitate comparison across energies, times and numbers of ICs considered. At very early times ($t\lesssim 10$), most cells have not been visited by trajectories, so that the histogram is strongly peaked around zero counts. As $t$ increases, we observe from Fig.~\ref{f:approach-to-mixing-in-time} that the histograms shift, and become progressively narrower, peaking around the expected value of 1 with the expected width (see Fig.~\ref{f:stdev-of-ntilde}). This provides evidence for mixing in the regime of global chaos. In Fig.~\ref{f:approach-to-mixing-in-energy}, we compare these histograms at sufficiently late times ($t = 300$) for a range of energies and observe significant departures from mixing for energies outside the band of global chaos. In fact, for energies such as $E = 4.5$ and $E = 6$, the histograms in Fig.~\ref{f:approach-to-mixing-in-energy} show three distinct peaks corresponding to cells that are never visited and two other types of cells (in chaotic regions) that are visited with unequal frequencies  (see Fig.~\ref{f:psec-outside-globalchaos}). This characteristic departure from mixing with respect to the Liouville measure (even when restricted to chaotic regions) is also reflected in the two distinct densities of points in Poincar\'e plots at such energies, as seen in Fig.~\ref{f:psec-outside-globalchaos}.

\section{Recurrence time statistics}
\label{s:recurrenc-time-dist}

\begin{figure*}	
	\centering
	\begin{subfigure}[t]{8cm}
		\centering
		\includegraphics[width=8cm]{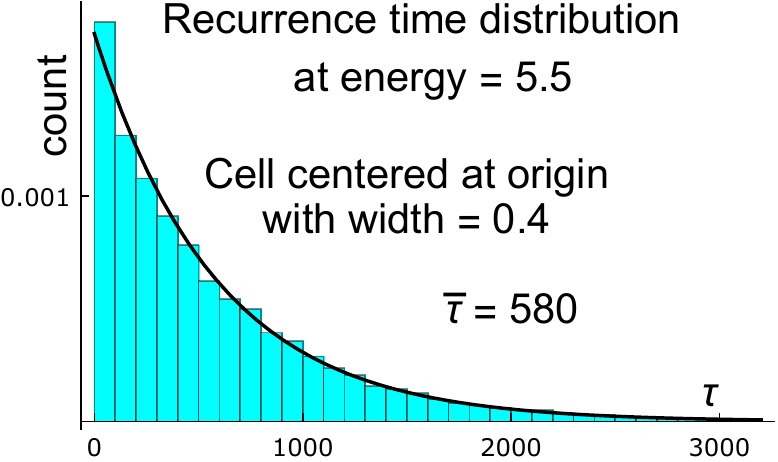}
		\caption{ \small   }
		\label{f:recurrence-time-distribution-a}
	\end{subfigure}
\quad
	\begin{subfigure}[t]{8cm}
		\centering
		\includegraphics[width=8cm]{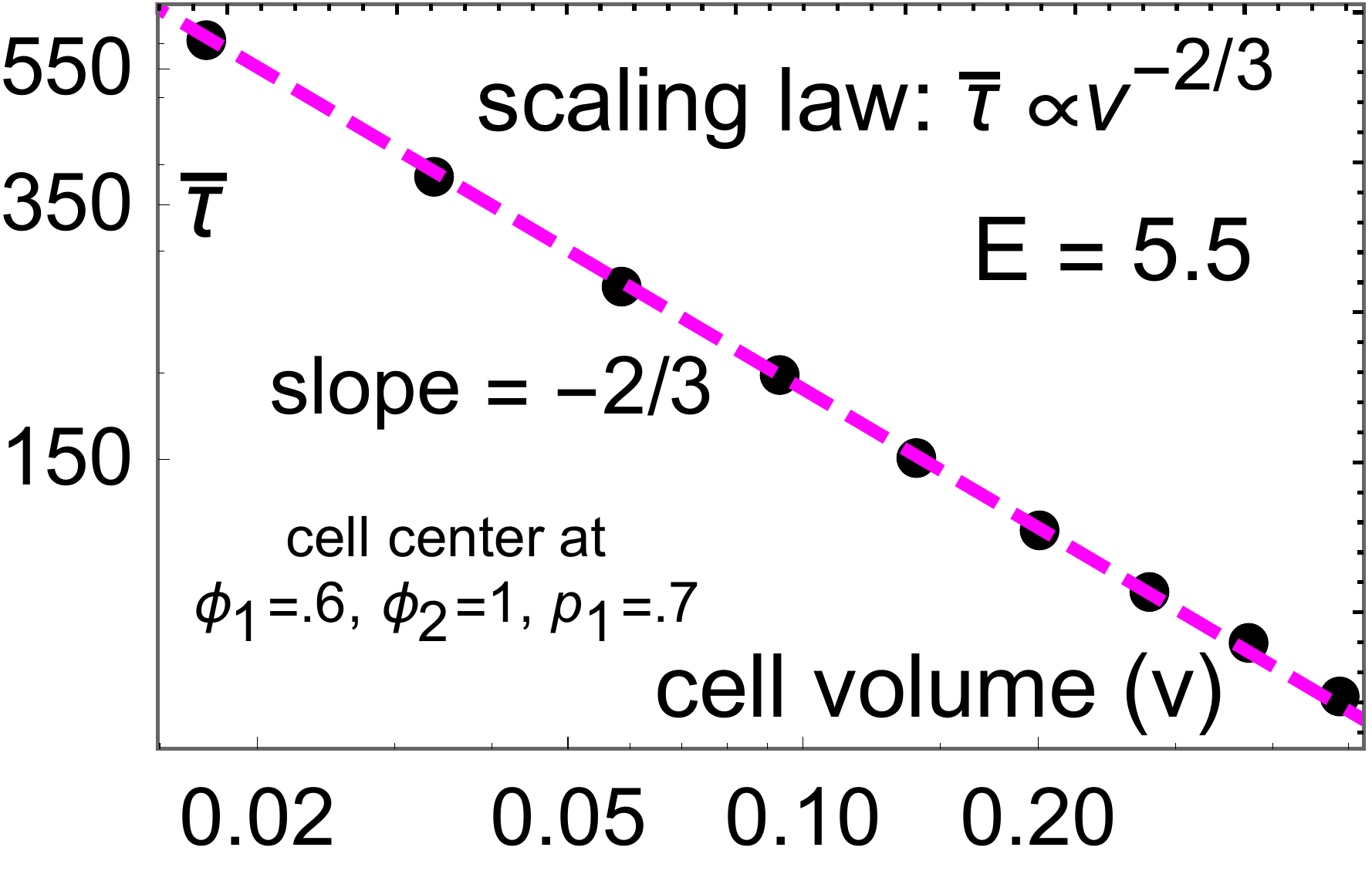}
		\caption{ \small   }
		\label{f:recurrence-time-distribution-b}
	\end{subfigure}
	\begin{subfigure}[t]{8cm}
		\centering
		\includegraphics[width=8cm]{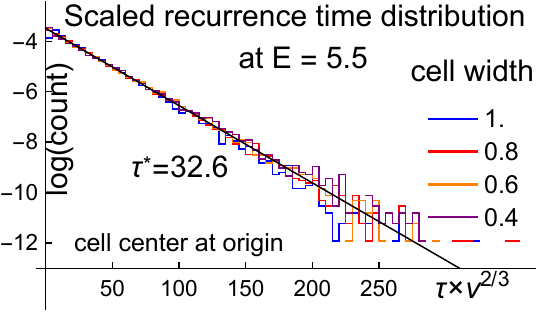}
		\caption{ \small   }
		\label{f:recurrence-time-distribution-c}
	\end{subfigure}
\quad
	\begin{subfigure}[t]{8cm}
		\centering
		\includegraphics[width=8cm]{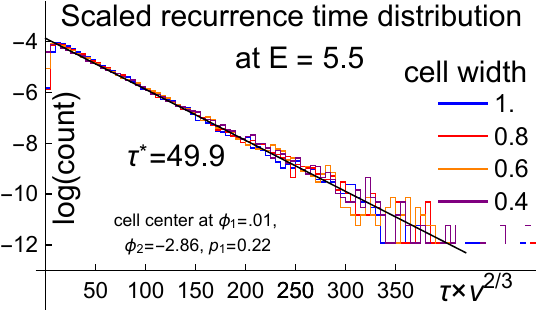}
		\caption{ \small   }
		\label{f:recurrence-time-distribution-d}
	\end{subfigure}
	\caption{ \small   (a) Histogram of recurrence times (normalized to be a probability distribution) for a cubical cell centered at the origin ($p_1=\vf_1=\vf_2= 0$) of the globally chaotic energy-$5.5$ (in units where $m = r = g = 1$) hypersurface showing an exponential law $(1/\bar\tau)\exp(-\tau/\bar\tau)$ where $\bar\tau$ is the fitted mean recurrence time. Note that $\bar \tau \approx 580$ is much larger than the time scale of the linearized system ($1/\om_0 =\sqrt{mr^2/3g}$). (b) At any cell location, $\bar \tau$ scales as the minus two-thirds power of the Liouville volume of the cell, consistent with ergodicity. (c, d) Normalized histogram of (recurrence times) $\times$ (cell volume)$^{2/3}$ plotted on a log-linear scale for cells of various widths, showing a universal exponential distribution $(1/\tau^*){\rm exp}(-\tau/\tau^*)$ away from very small $\tau$. The larger spread at large $\tau \times {\rm v}^{2/3}$ is due to lower statistics. The rescaled fitted mean recurrence time $\tau^*$ varies with cell location but only weakly depends on energy within the band of global chaos.}
	\label{f:recurrence-time-distribution}
\end{figure*}

Here, we study the statistics of Poincar\'e recurrence times to a three-dimensional cell in an energy-$E$ hypersurface of the phase space. For convenience, we choose the cell to be a cuboid of width $w$, e.g, $-w/2 \leq \vf_1, \vf_2, p_1 \leq w/2$ with $p_2 = p_2^+$ (\ref{e:p2PM}) determined by energy for a cell centered at the origin. We choose a large number ($\sim 3 \times 10^4$) of initial conditions distributed uniformly randomly within the cell and numerically evolve them forward in time. The recurrence time $\tau$ for a given trajectory is defined as the time from the first exit to the next exit from the cell \cite{zaslavsky}. Evidently, starting from the instant the trajectory first exits the cell, $\tau$ is the sum of the times it spends outside the cell and while traversing the cell. A histogram of the recurrence times (normalized to be a probability distribution) is then plotted as in Fig. \ref{f:recurrence-time-distribution-a}. 

{\fl \bf Exponential law:} For uniformly mixing dynamics, it is expected that this normalized distribution follows an exponential law $(1/\bar \tau) e^{-\tau/\bar\tau}$ where $\bar \tau$ is the mean recurrence or relaxation time \cite{zaslavsky}. As shown in Fig. \ref{f:recurrence-time-distribution}, this exponential law for recurrence times holds for energies in the band of global chaos though there can be (sometimes significant) deviations for very small values of $\tau$ (e.g., $\tau \lesssim 25 \ll \bar \tau \approx 250$ for $w$ = .6 in Fig.~\ref{f:recurrence-time-distribution-d}). These deviations could be attributed to a memory effect, the finite time that the system takes before the dynamics displays mixing (see Fig. \ref{f:approach-to-mixing-in-time}). Thus, $\bar \tau$ is to be interpreted as the time constant in the above exponential law that best fits the distribution away from very small $\tau$.

A heuristic argument for the exponential law follows; for a more detailed treatment, see Refs.~\cite{kac, balakrishnan-pre} and references therein. We pick a large number $N$ of ICs uniformly from a region $\Om$ of volume $V_\Om$ in an energy-$E$ hypersurface of volume $V_E$. They are evolved in time and their locations sampled at a temporal frequency $\Delta$. At each such instant, the probability of returning to $\Om$ is $p = V_\Om/V_E$ provided a sufficiently long time $T$ has elapsed for correlations to have died out. Suppose a fraction $f$ of trajectories have {\it not} returned to $\Om$ by this time $T$. Then, the probability that the {\it first return time} $\tau$ equals $T + \Delta$ is $P(\tau = T + \Delta) = f p$ (leaving aside possible returns that the sampling at frequency $\Delta$ does not detect). If $\Delta$ is chosen large enough ($\gtrsim$ transit time across $\Om$), we also have $P(\tau = T + 2\Delta) = f (1-p) p$ and similarly $P(\tau = T + n \Delta) = f (1-p)^{n-1} p$ for $n = 1,2,\cdots$. In the limit $N \to \infty$, $\Delta \to 0$ and $V_\Om \to 0$ holding $\Delta/p = \bar \tau$ fixed, and omitting prefactors (independent of $t$) that go into the normalization,
		\beq
		P( t \leq \tau \leq t+dt) \propto \lim_{\Delta \to 0} (1 - p)^{t/\Delta} = e^{-t/\bar \tau}.
		\eeq

{\fl \bf Scale invariance:} Though $\bar\tau$ varies with the width $w$, we find that when rescaled by the two-third power of the Liouville volume ${\rm v}$ of the cell, it becomes independent of cell size within the band of global chaos. In other words, $\bar\tau \times {\rm v}^{2/3} = \tau^*$ is constant for cells centered at a given location (see Fig. \ref{f:recurrence-time-distribution-b}). Thus, as shown in Figs.  \ref{f:recurrence-time-distribution-c} and \ref{f:recurrence-time-distribution-d}, the rescaled recurrence time distributions for various cell sizes, all follow the {\it same} exponential law for a given energy and cell center. This scaling law may be viewed as a 3d energy hypersurface analogue of the 2d phase space version given in Eq. (36) of Ref.~\cite{balakrishnan-scaling-law} as well as of the scaling law for the mean recurrence time of the second type in Ref.~\cite{gao}. Heuristically, the mean recurrence time $\bar \tau$ is inversely proportional to the surface area ($\sim {\rm v}^{2/3}$) of the cell and allows us to view the `attractor' as being three dimensional, which is consistent with global chaos and ergodicity. On the other hand, we find that the scaling exponent deviates from two-thirds in chaotic regions outside this band. This is to be expected since the dynamics at such energies is not mixing in such chaotic regions, as shown in Figs.~\ref{f:approach-to-mixing-in-energy} and \ref{f:psec-outside-globalchaos}.

The above scaling law defines for us the  {\it scaled} mean recurrence time $\tau^*$ for cells centered at a given location of an energy hypersurface. We find that $\tau^*$ varies with location. For instance, for cells centered along an isosceles trajectory (see \S \ref{s:formulation}), we find that the values of $\tau^*$ display a reflection symmetry about the triple collision configuration and vary over the range $31 \lesssim \tau^* \lesssim 56$. On the other hand, within the band of global chaos, $\tau^*$ hardly varies with energy for a given cell location.

{\bf \fl Loss of memory:} We also observe the absence of memory in the sense that the gaps between successive recurrence times are uncorrelated. For instance, let us denote by $\tau_1$ and $\tau_2$ the first recurrence time and the gap between second and first recurrence times for a given trajectory and cell, and define the the correlation coefficient
	\beq
	\scripty{r} = 
	\left[ { \bra \tau_1 \tau_2 \ket - \bra \tau_1 \ket \bra \tau_2 \ket} \right] / ({\sig_1 \sig_2}).
	\eeq
The averages here are performed with respect to a random collection of trajectories and $\sig_{1,2}$ denote the standard deviations of $\tau_{1,2}$. We find that $|\scripty{r}| \approx 10^{-3}-10^{-5} \ll 1$ for cells of widths $0.4-1.2$ centered at the origin of the energy $E = 5.5$ hypersurface, indicating uncorrelated recurrences.

\section{Discussion}
\label{s:discussion}

In this paper on the classical three rotor problem, we have provided evidence for ergodicity and mixing in a band of energies where the dynamics is globally chaotic. \S \ref{s:formulation} contains a summary of the main results. Here, we mention a few open questions arising from this work. While we have provided a qualitative explanation for the shape of the momentum distribution over energy hypersurfaces in \S \ref{s:ergodicity-time-ensemble-avgs}, it would be nice to understand the mechanisms underlying the phase transitions observed in $\rho(p_1)$. In another direction, outside the band of global chaos, it would be interesting to determine whether the dynamics, when restricted to a chaotic region, is ergodic and/or mixing with respect to a suitable measure. In fact, Figs. \ref{f:approach-to-mixing-in-energy} and \ref{f:psec-outside-globalchaos} suggest that this measure cannot be the Liouville measure. In \S \ref{s:recurrenc-time-dist}, the scaled mean recurrence time $\tau^*$ to cells at a given location was found to vary with the location on the energy hypersurface. It would be of interest to study the nature of this variation and its physical implications. We also wonder whether global chaos and ergodicity are to be found in the problems of four or more rotors.

Unlike billiards and kicked rotors, the equations of the three-rotor system do not involve impulses/singularities. It would be interesting to identify other such continuous time autonomous Hamiltonian systems that display global chaos and ergodicity. As noted, the three rotor problem may also be formulated as geodesic flow on a manifold of non-constant Jacobi-Maupertuis curvature. A challenging problem would be to try to extend the analytic treatments of ergodicity in geodesic flows on constant curvature Riemann surfaces to the three rotor problem. Finally, an exploration of ergodicity and recurrence in the quantum three rotor  system is desirable, given that it models chains of coupled Josephson junctions.

\begin{acknowledgments} 

We thank M Berry, A Lakshminarayan, R Nityananda and anonymous reviewers for helpful comments and references. This work was supported in part by the Infosys Foundation, J N Tata trust and grants (MTR/2018/000734, CRG/2018/002040) from the Science and Engineering Research Board, Govt. of India.

\end{acknowledgments}

\appendix
\section{Power-law approach to ergodicity in time}
\label{a:approach-to-ergodicity}

Assuming correlations decay sufficiently fast, as expected for a chaotic system, we give here a heuristic explanation for our observed (see \S \ref{s:approach-to-ergodicity}) power-law approach to ergodicity in time (see also Ref.\cite{prl-dechant} for a discussion based on a stochastic framework). Let $F(p, \vf)$ be a dynamical variable with ensemble average at energy $E$ denoted $\bar F = \bra F \ket_{\rm e}$  (\ref{e:ensemble-avg-distr-vol-egy-surf}). Its time average, over the interval $[0,T]$, along an energy-$E$ phase trajectory $( \vec p_i(t), \vec \vf_i(t))$ labelled $i$, is denoted
	\beq
	\tilde F_{i}(T) = \ov{T} \int_0^T F_i(t) \: dt \equiv \ov{T} \int_0^T F(\vec p_i(t), \vec \vf_i(t)) \: dt.
	\eeq
To examine the rate at which time averages along different trajectories $i$ approach the ensemble average, we define the mean square deviation of $\tilde F_{i}(T)$ from $\bar F$ for a family $\cal I$ of trajectories:
	\beq
	{\rm var}_F(T) = \Bra \left( \tilde F_{i}(T) - \bar F \right)^2 \Ket \equiv \ov{\# (\cal I)} \sum_{i \in \cal I} \left( \tilde F_{i}(T) - \bar F \right)^2.
	\eeq
Expanding, we write the mean square deviation as
	\beq
	{\rm var}_F(T) = \Bra \tilde F_{i}(T)^2 \Ket  + \bar F^2 - 2 \bar F \Bra \tilde F_{i}(T) \Ket.
	\eeq
We now assume that the ICs for the trajectories in ${\cal I}$ are distributed uniformly with respect to the Liouville measure on the energy-$E$ hypersurface. Since the dynamics is Hamiltonian, by Liouville's theorem the trajectories remain uniformly distributed at all times $T$, so that as ${\# (\cal I)} \to \infty$,
	\beq
	\Bra \tilde F_{i}(T) \Ket = \bar F.
	\eeq
Thus, the mean square deviation becomes
	\beqs
	{\rm var}_F(T) &=& \Bra \tilde F_{i}(T)^2 \Ket - \bar F^2 =  \Bra \tilde F_{i}(T)^2 - \bar F^2 \Ket \cr
	&=& \Bra \ov{T^2} \int_0^T \int_0^T  [F_i(t_1) F_i(t_2) - \bar F^2] dt_1 dt_2 \Ket \cr
	&=& \ov{T^2} \int_0^T \int_0^T  \Bra F_i(t_1) F_i(t_2) - \bar F^2 \Ket
 dt_1 dt_2. 	\quad
 	\eeqs
We now assume that $F_i(t_1)$ and $F_i(t_2)$ are practically uncorrelated if $|t_1 - t_2| > \eps$ for some time $\eps$, i.e.,
	\beq
	\Bra F(t_1) F(t_2) - \bar F^2 \Ket \approx \begin{cases}0 & {\rm if} \quad |t_1 - t_2| > \eps  \quad \text{and} \\
	{\cal C}(t_1-t_2) & {\rm otherwise} \end{cases}
	\eeq
by time-translation invariance, for some (2nd cumulant) function ${\cal C}(t_1-t_2)$. We now change integration variables from $t_{1,2}$ to $u = t_1 - t_2$ and $v = (t_1+t_2)/2$ with $dt_1 dt_2 = du \: dv$ and assume $T \gg \eps$ to get
	\beqs
	{\rm var}_F(T) 
	&\approx& \ov{T^2} \int_{0}^T dv \int_{-\eps}^\eps  du \: {\cal C}(u)    
	= \ov{T}  \int_{-\eps}^\eps  {\cal C}(u) du.  \qquad
	\eeqs
Thus, the RMS deviation of time averages from the ensemble average vanishes like $1/\sqrt{T}$ as $T \to \infty$.

\footnotesize



\begin{thebibliography}{99}

\bibitem{sinai-geodesic-flow} Ya. G. Sinai, {\it Geodesic flows on manifolds of negative constant curvature}, Dokl. Akad. Nauk SSSR, {\bf 131}(4), 752 (1960) [English translation: Sov. Math. Dokl., {\bf 1}, 335 (1960)].

\bibitem{sinai-central-limit} Ya. G. Sinai, {\it The central limit theorem for geodesic flows on manifolds of constant negative curvature}, Dokl. Akad. Nauk SSSR, {\bf 133}(6), 1303 (1960) [English translation: Sov. Math. Dokl., {\bf 1}, 983 (1960)].

\bibitem{sinai-billiard} Ya. G. Sinai, {\it Dynamical Systems with Elastic Reflections}, Russ. Math. Surv., {\bf 25},  137 (1970).

\bibitem{lenci} M. Lenci, {\it Aperiodic Lorentz gas: recurrence and ergodicity}, Ergodic Theory Dynam. Systems, {\bf 23}(3), 869 (2003).

\bibitem{chirikov} B.~V.~Chirikov, {\it A universal instability of many-dimensional oscillator systems}, Phys. Rep., {\bf 52}, 263 (1979).

\bibitem{stdmap-ergodicity} A. Giorgilli and V. F. Lazutkin, {\it Some remarks on the problem of ergodicity of the Standard Map}, Phys. Lett. A, {\bf 272}(5-6), 359 (2000).

\bibitem{gskhs-cnsd-3rotor} G. S. Krishnaswami and H. Senapati, {\it Stability and chaos in the classical three rotor problem}, Indian Academy of Sciences Conference Series, {\bf 2}(1), 139 (2019); \href{https://arxiv.org/abs/1810.01317}{arXiv:1810.01317}.

\bibitem{gskhs-3rotor} G. S. Krishnaswami and H. Senapati, {\it Classical three rotor problem: periodic solutions, stability and chaos}, Chaos, {\bf 29}(12), 123121 (2019); \href{https://arxiv.org/abs/1811.05807}{arXiv:1811.05807}.

\bibitem{Rabouw-Ruijgrok} F. Rabouw and Th. W. Ruijgrok, {\it Three particles on a ring}, Physica A, {\bf 109}(3), 500 (1981).

\bibitem{sondhi-girvin} S. L. Sondhi, S. M. Girvin, J. P. Carini and D. Shahar,  {\it Continuous quantum phase transitions}, Rev. Mod. Phys., {\bf 69}(1), 315 (1997).

\bibitem{kac} M. Kac, {\it Probability and Related Topics in Physical Sciences}, Interscience, London (1959).

\bibitem{zaslavsky} G. M. Zaslavsky, {\it Chaos, fractional kinetics, and anomalous transport}, Phys. Rept., {\bf 371}, 461 (2002).

\bibitem{hirata-axiomA} M. Hirata, {\it Poisson law for Axiom A diffeomorphisms}, Ergodic Theory Dynam. Systems, {\bf 13}(3), 533 (1993).

\bibitem{hirata-uniformly-hyperbolic} M. Hirata, {\it Dynamical Systems and Chaos}, Vol. 1, p. 87, World Scientific, New Jersey (1995).

\bibitem{gskhs-3body} G. S. Krishnaswami  and H. Senapati, {\it Curvature and geodesic instabilities in a geometrical approach to the planar three-body problem}, J. Math. Phys., {\bf 57}, 102901 (2016).

\bibitem{prl-dechant} A. Dechant, E. Lutz, D. A. Kessler and E. Barkai, {\it Fluctuations of Time Averages for Langevin Dynamics in a Binding Force Field}, Phys. Rev. Lett., {\bf 107}, 240603 (2011).

\bibitem{altmann} E. G. Altmann, E. C. da Silva and I. L. Caldas, {\it Recurrence Time Statistics for Finite Size Intervals}, Chaos, {\bf 14}(4), 975 (2004).

\bibitem{balakrishnan-scaling-law} C. Nicolis, G. Nicolis, V. Balakrishnan and M. Theunissen, {\it Recurrence time statistics in low-dimensional dynamical systems}, in Stochastic Dynamics, Eds. L. Schimansky-Geier and T.  P\"oschel, Springer-Verlag, Berlin (1997).

\bibitem{gao} J. B. Gao, {\it Recurrence Time Statistics for Chaotic Systems and Their Applications}, Phys. Rev. Lett., {\bf 83}(16), 3178 (1999).


\bibitem{arnold-avez} V. I. Arnold and A. Avez, {\it Ergodic Problems of Classical Mechanics}, W. A. Benjamin, New York (1968).

\bibitem{gutzwiller-book} M. C. Gutzwiller, {\it Chaos in Classical and Quantum mechanics}, Springer-Verlag, New York (1990).

\bibitem{balakrishnan-pre} V. Balakrishnan, G. Nicolis and C. Nicolis, {\it Recurrence time statistics in deterministic and stochastic dynamical systems in continuous time: A comparison}, Phys. Rev. E, {\bf 61}(3), 2490 (2000). 

\end{thebibliography}
\end{document}